\def\aj{{AJ}}

\def\apj{{ApJ}}
\def\apjs{{ApJS}}

\def\mnras{{MNRAS}}

\documentclass[cup5b]{caps}
\usepackage{graphicx}
\usepackage{amssymb}
\usepackage{ociwsymp1e}
\HeadText{I. Strateva, M. A. Strauss, and L. Hao}

\begin{document}

\pagenumbering{arabic}

\author[]{I. STRATEVA, M. A. STRAUSS, and L. HAO
\\Princeton University}

\chapter{Double-Peaked Broad Emission Lines \\ and the Geometry of Accretion in 
AGNs}

\begin{abstract}

Although accretion disks are a theoretically appealing model for the
geometry and dynamics of the gas in the vicinity of black holes in
Active Galactic Nuclei (AGN), there is little direct observational
evidence for their existence. The telltale signature of disk emission
in AGN -- \emph{double peaked emission lines} -- have so far been
found in only two dozen cases in the optical ([1], [2]). We have selected
$\sim$ 100 double-peaked broad emission line AGN from the Sloan
Digital Sky Survey (SDSS) from a large sample of over 4000 AGN with
$z<0.4$. By comparing the properties of these AGN with those of the
full sample, we hope to isolate the defining characteristics of
disk-emitters and ultimately answer the question: ``If all AGN have
accretion disks, why don't they all show double-peaked disk  emission
lines?'' Here we present Gaussian  parameterized H$\alpha$
line-profile measurements for the sample of double-peaked  AGN in
comparison with circular and elliptical accretion disk models.

\end{abstract}

\section{Introduction}
Disk accretion provides a highly efficient mechanism for the
dissipation  of angular momentum, allowing for the conversion of large
fraction of the rest-mass energy of the material accreted onto a black
hole in the form of radiation. Because AGN are the most luminous
compact sources observed, accretion disks are believed to be
responsible for their continuum  emission and the formation of jets.
Mass-transferring binary systems, containing stellar mass black holes
(\emph{e.g.} low mass X-ray binaries) or other compact objects
(\emph{e.g.} white dwarfs in cataclysmic variables and dwarf novae),
provide numerous examples of disk accretion. Extensive light-curve and
time-lag (both continuum and emission-line) observations of these binary
systems give conclusive evidence for the existence of accretion disks
([3], [4]), along with  detailed temperature and surface density maps
and striking examples of hot spots and spiral structures (\emph{e.g.}
IP Peg, [5]).  The emission signatures of disks in AGN are much more
elusive, owing to the impossibility of resolving the disk structures
and following their time variation in even the nearest galactic
nuclei.  The complex interplay of disk radiation reprocessing in hot
coronas, jets and winds, driven by the presence of multiphase plasma
in the vicinity of the central engine, make it virtually impossible to
compute the AGN spectral energy distribution from first principles,
further frustrating the interpretation of observations.

Despite the complexity of the black hole environment, a small class of
AGN do show the double-peaked broad lines characteristic of disk
emission in the optical. These AGN present us with a unique
opportunity  to study the structure of AGN accretion disks.

\section{Data Analysis}
The Sloan Digital Sky Survey ([6], [7]) will produce upon
completion a catalog of over $10^5$ multicolor selected AGN with
five-band  (\emph{ugriz}) photometry and calibrated spectra covering
the wavelength region 3800--9200\AA  ~at a spectral resolution  of
1850--2200. The identification and basic measurements for the sample
are done automatically by a series of custom pipelines. The
spectroscopic  observations are carried out using the $2.5$m SDSS
telescope at Apache Point Observatory and a pair of double, fiber-fed
spectrographs.

\begin{figure}
\centering
\includegraphics[width=\columnwidth]{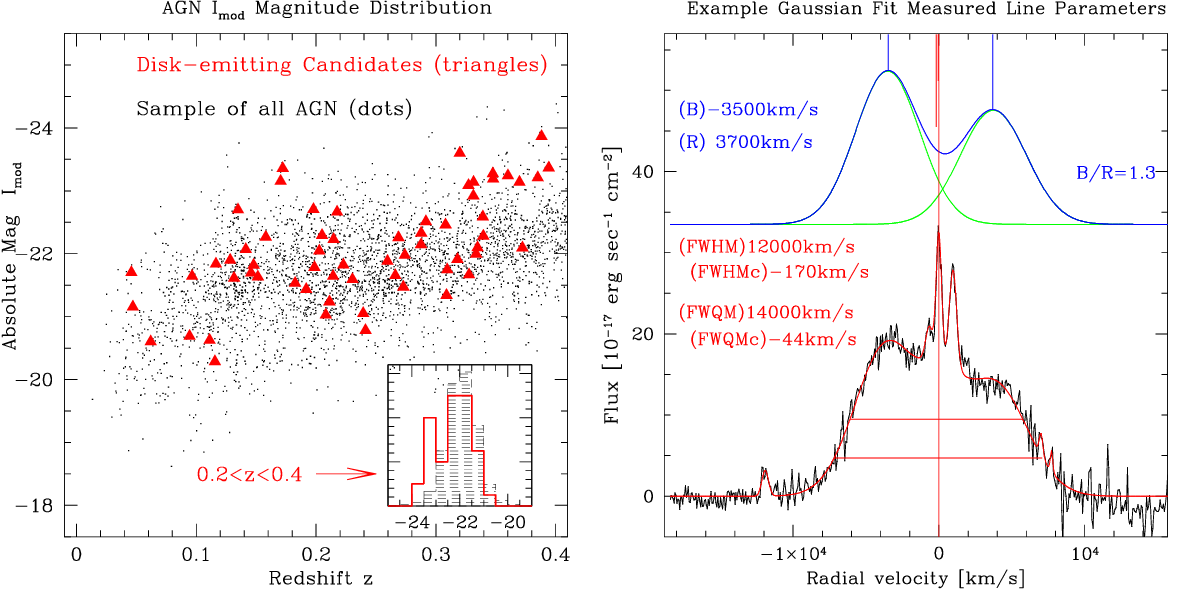}
\caption{ \emph{Left:} Absolute magnitude distribution of
candidate disk-emitters (red)  compared to that of the full AGN sample
over the  same redshift range (black).  
\emph{Right:} Example Gaussian fit. The original profile is in black, 
the Gaussian fit overlaid in red. The blue line (displaced vertically) 
shows the fitted double-peaked H$\alpha$ after subtraction of the narrow 
lines and a central broad H$\alpha$ component.}
\label{absmag}
\end{figure}

Using the SDSS spectroscopic survey we selected $\sim$100
double-peaked  broad line AGN based on their H$\alpha$ line profiles.
The absolute magnitude distribution of the sample vs. redshift is
presented in the left panel of Figure \ref{absmag} in comparison with
that of all AGN.  In order to characterize  the line profiles in a
model-independent  way, we subtract a sum of stellar  + power law
continua ([8]) and fit a sum of Gaussians to each H$\alpha$ line
complex (see the right panel of Figure \ref{absmag} for an example).
We use the resulting smooth Gaussian profiles to estimate the
positions (in velocity space relative to the narrow H$\alpha$
component) of the blue and red peaks, the blue-to-red peak ratio, the
peak separation, the FWHM and the FW$\frac{1}{4}$M (\emph{i.e.},
measured at $\frac{1}{4}$ maximum) of the broad  line component and
the displacements of their respective centroids.  By comparing the
distribution of these line-characterizing quantities with those
measured on a set of accretion disk models, we find the distribution
of disk model parameters that best describe the data in a statistical
sense. For the present comparison we have chosen one axisymmetric
(circular) accretion disk model with intrinsic turbulent broadening
[9] and the simplest non-axisymmetric disk model, that of an
elliptical disk [10]. 

\section{Accretion Disk Models}
We created \hbox{$\sim$25,000} model disk-emission line profiles for
comparison with observations, by varying all model parameters on a
grid. The circular disk model (after [9]) assumes a simple
relativistic Keplerian disk which is geometrically thin and optically
thick and has five parameters: the inner and outer radii of the
emitting ring (in units of the gravitational radius,
$R_G=2GM_{\bullet}/c^2$), $\xi_1$ and $\xi_2$, the disk inclination,
$i$, the slope of the surface emissivity power law, $q$, and the
turbulent broadening, $\sigma$. These parameters correspond to the
five parameters  of a two-Gaussian fit representation of the
double-peaked profile,  \emph{i.e.}, the model is fully constrained by
the Gaussian fit.  Doppler boosting of radiation results in a higher
blue peak than red in the circular disk case, and a net redshift of
the whole line is observed.

\begin{figure}
\centering
\includegraphics[width=\columnwidth]{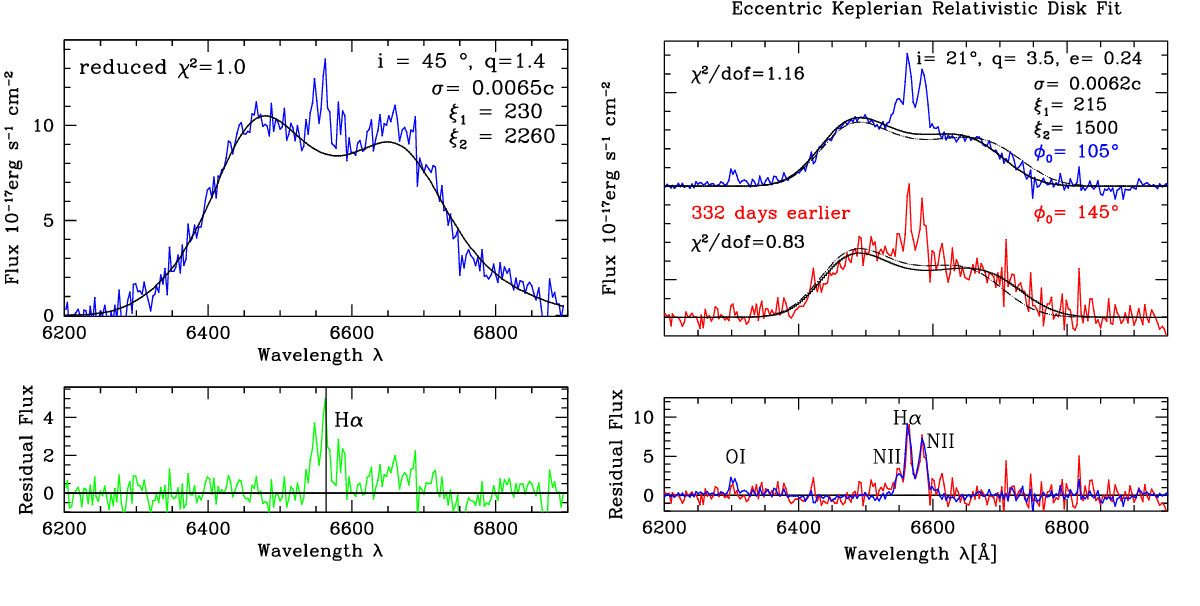}
\caption{\emph{Left:} circular disk fit and model parameters
(top)  and residual narrow lines (bottom).  \emph{Right:}
elliptical disk fits at two epochs (the red spectrum was observed
one year before the blue; top) and narrow line residuals (bottom).
The observed disk precession of $\sim 40^\circ$ between the two
epochs,  implies a precession period of just over 8 years. The
precession period,  $P_{GR}$, is approximately $P_{GR} \approx 2
\times M_7(\xi/200)^{2.5}$years ([1]), leaving us with a  plausible
black hole mass of $\sim 2.5\times10^6M_{\odot}$. This is, of course,
an over-interpretation of 2 epochs of data; the example is merely
intended to be illustrative.}
\label{fits}
\end{figure}

If the red peak is stronger than the blue, or the profile is  observed
to be variable in this sense, the circular disk emission profile model
fails and some asymmetry in the disk has to be invoked to reproduce
the line asymmetry. Common choices are elliptical disks (thought to
arise  when a single star is disrupted in the vicinity of a black
hole, [10]), warped disks (theorized to exist around rotating  black
holes, [11]; spiral disks ([12], [13]), or disks with a
hot spot ([14]). The choice of the elliptical disk to represent  all
non-axisymmetric disks in this comparison is justified solely by its
relative simplicity and the fact that we do not have the extensive
time variability observations required to distinguish between the
different non-axisymmetric models.  Example circular and elliptical
disk fits are given in Figure \ref{fits}. Note that a
\emph{statistical}  comparison of the observed and disk-model line
profiles, as the one we  present here, is more meaningful in terms of
characterizing the  distribution of accretion disk parameters in AGN
than individual  kinematic profile fits, especially in view of the
observed frequent  variability of the line profiles. Individual model
fits will be presented in a follow up paper.

\begin{figure}
\centering
\includegraphics[width=\columnwidth]{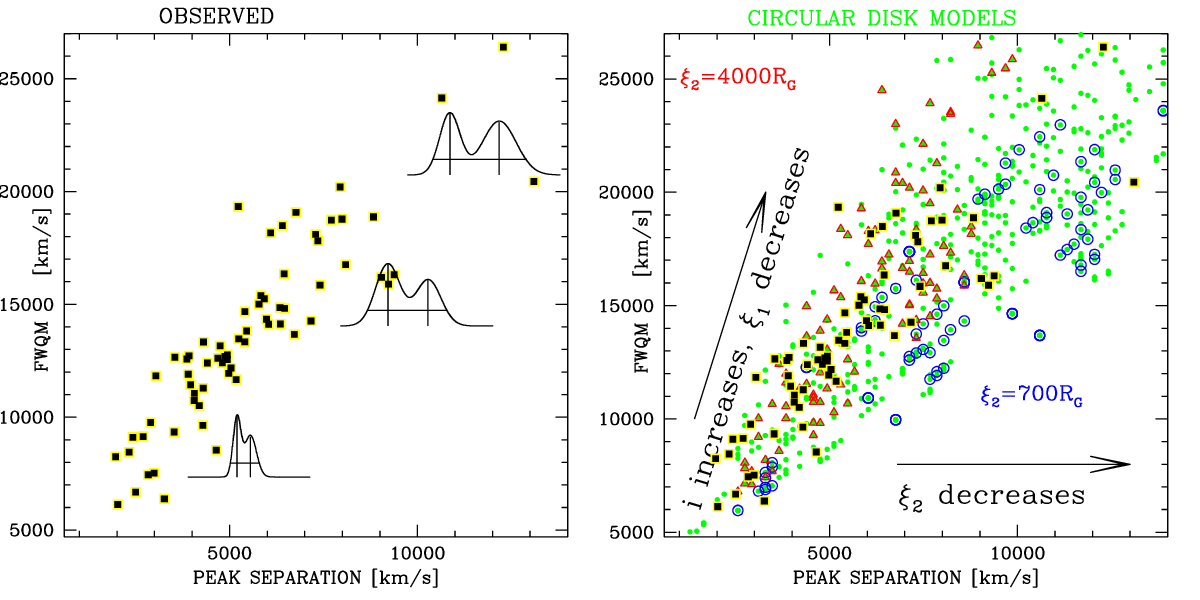}
\caption{\emph{Left:} {\bf Observed} peak separation
vs. FW$\frac{1}{4}$M and 3 Gaussian fits for illustration.
\emph{Right:} {\bf Circular disk models.} Each point corresponds to a
double peak model realization. The inclination $i$ (larger effect) and
inner radius, $\xi_1$,  determine the FW$\frac{1}{4}$M, while the peak
separation is governed by the outer radius,  $\xi_2$. The red
triangles (blue circles) correspond to models with largest  (smallest)
outer radii.}
\label{circular}
\end{figure}

\section{Statistical Comparison of Disk-Model and Observed Line Profiles}
\begin{figure}
\centering
\includegraphics[width=\columnwidth]{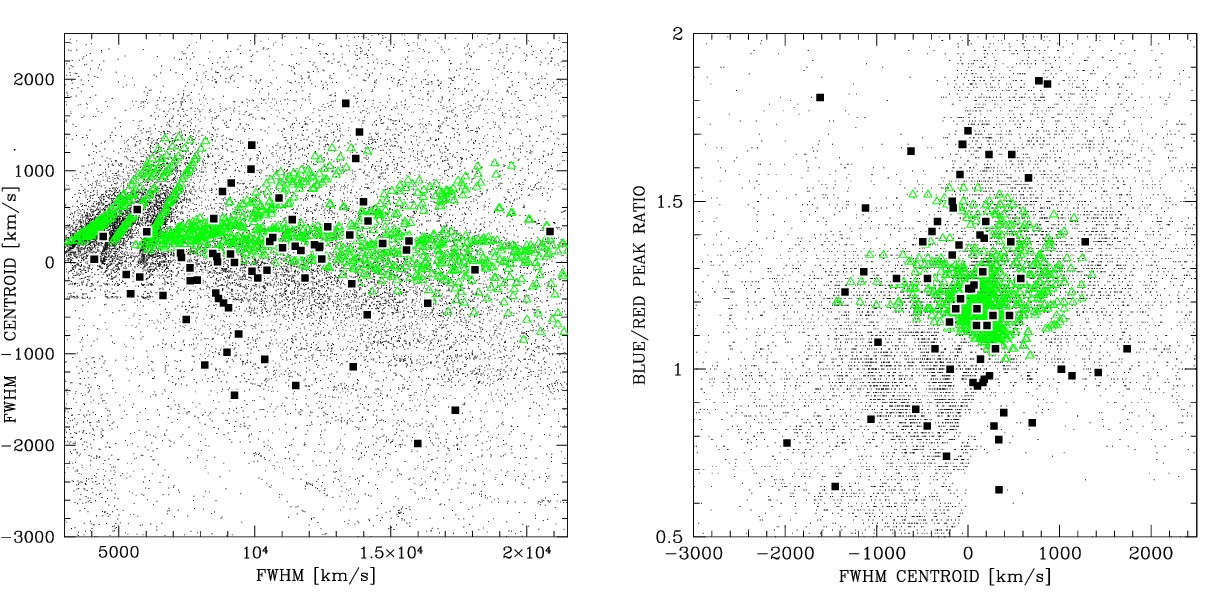}
\caption{ {\bf Observed vs. model comparison suggests the need for
non-axisymmetric disks.}  Black squares stand for observed quantities,
green triangles for axisymmetric (circular) disk models, and black
dots for non-axisymmetric (elliptical) models.  \emph{Left:} FWHM
vs. FWHM centroid. \emph{Right:} Peaks centroid vs. Blue/Red  peak
height ratio.}
\label{asymmetric}
\end{figure}

The left panel of Figure \ref{circular} presents the observed peak
separation \emph{vs.}\ the FW$\frac{1}{4}$M of the emission line which
we compare to a select subset of circular disk models in the right
panel. The FW$\frac{1}{4}$M of a model line profile increases most strongly 
with increasing disk inclination, $i$, and also with decreasing
inner radius of the emitting region, $\xi_1$, while the peak
separation increases with decreasing outer radius, $\xi_2$. The
current sample of disk emitters is consistent with accretion disks
with inclinations $15^{\circ}<i<45^{\circ}$, inner radii of
$200R_G<\xi_1<600R_G$ and outer radii of $1500R_G<\xi_2<4000R_G$.
Emission lines produced in disks of smaller inclinations are
single-peaked, and thus not selected from the observed data for this
comparison. There is no \emph{a priori} reason why we should not
observe disks at higher inclinations, so the lack of such observed
profiles could be telling us that we are looking through the dusty
torus believed to exist at larger radii in unified models of AGN. By
comparing the position of the FWHM centroid of the observed line
profiles and the disk models, we conclude that at least half of  the
cases require non-axisymmetric disks.  The results of this  comparison
are illustrated in the right panel of Figure \ref{asymmetric}.

\section{Conclusions}
Further exploration of the physical conditions of the disk through
kinematic modeling of the individual line profiles,  combined with
flux measurements in different wavebands, and  complemented by
variability and spectropolarimetric observations will eventually  help
us sketch the geometry of accretion in these objects, while comparison
of the properties of these select AGN with the rest of the sample
allows us to address the question of the lack of obvious disk emission
in the full AGN sample. Such statistical studies were impossible on
the basis of the small samples existing before.  The current SDSS
sample is $\sim$4 times larger than previous samples (\emph{e.g.},
[1], [2]) and will increase twofold by the survey end.

{\bf Acknowledgments.}
We would like to thank Michael Eracleous, Pat Hall, Li-Xin Li, Wei
Zheng, and Bohdan Paczynski for their guidance and useful discussions. 
Funding for the SDSS is provided by the Alfred P. Sloan Foundation,
NASA, NSF, DoE, Monbukagakusho, the Max Planck Society and the member
institutions.  

\end{document}